# Quantum Computing and Nuclear Magnetic Resonance

**Jonathan A. Jones**[a]

[a] *Oxford Centre for Quantum Computation, Clarendon Laboratory, Parks Road, Oxford OX1 3PU, UK.*
*Fax: +44 1865 272387; Tel: +44 1865 272247; E-mail: jonathan.jones@qubit.org*

Quantum information processing is the use of inherently quantum mechanical phenomena to perform information processing tasks that cannot be achieved using conventional classical information technologies. One famous example is quantum computing, which would permit calculations to be performed that are beyond the reach of any conceivable conventional computer. Initially it appeared that actually building a quantum computer would be extremely difficult, but in the last few years there has been an explosion of interest in the use of techniques adapted from conventional liquid state nuclear magnetic resonance (NMR) experiments to build small quantum computers. After a brief introduction to quantum computing I will review the current state of the art, describe some of the topics of current interest, and assess the long term contribution of NMR studies to the eventual implementation of practical quantum computers capable of solving real computational problems.

## Introduction

Information processing technologies, such as communication and computation technologies, are among the most characteristic features of the last few decades, and have revolutionised many areas of everyday life. Computing[1], in particular, has made extraordinary progress and seems to be advancing at breakneck speed: it is often noted that there is never a right time to buy a new computer, as an even better model will soon be on offer. Despite this apparently inexorable progress, however, current computing technology is rapidly approaching some fundamental limits.

These limits arise from the fact that the individual components on computer chips are rapidly approaching the atomic scale, at which point the laws of classical physics no longer apply. One solution to this problem is to use components that operate in an explicitly quantum mechanical fashion. A computer of this kind could, however, provide much more than a method for tackling the atomic limit: it could also use such essentially quantum mechanical properties as superposition and entanglement to perform information processing tasks beyond the reach of classical devices[2,3].

The theory of quantum information processing has been studied for many years[*], and some areas are now fairly well understood[4]. This has inevitably led to interest in devising practical quantum information technologies[5,6], capable of implementing quantum methods, but this is an extremely challenging task, and initial progress was quite slow. In 1996, however, David Cory and co-workers showed how techniques adapted from conventional liquid state NMR spectroscopy could be used to construct small quantum computing devices[7]. The first implementations of quantum algorithms[8–11] were described in 1998, and since then progress has been rapid[12], with NMR techniques lying far ahead of all other suggested quantum computing technologies.

This rapid progress has generated significant controversy: some authors have made much of the fact that current implementations of NMR quantum computing cannot be scaled up to produce devices capable of solving real problems, while others have claimed that NMR quantum computers are not in fact real quantum computers at all! While these claims have some merit they must be considered with caution.

### Moore's Law and computational complexity

The extraordinary progress in the development of computer technologies is usually summarised in the form of "Moore's Law". Put simply, Moore's Law[†] observes that the computer power available for a given sum of money roughly doubles every eighteen months. Like any form of exponential growth this doubling soon leads to enormous numbers: computer power increases tenfold every five years, one hundred fold every decade, and so on. Moore's Law has now held true for about forty years, during which time the power of computers has increased by a factor of about one hundred million.

Progress in the design of computers shows no sign of slowing down, and it is tempting to assume that Moore's Law can continue for many more years. In fact, it seems unlikely that it will be possible to continue for much more than another decade. The huge increases in computing power require corresponding decreases in the size of electronic components, and the limits of this approach, imposed by the atomic scale, are within sight[13]. For example, computer logic gates are built out of silicon MOSFET devices which depend on silicon dioxide dielectric layers, and when the thickness of these layers is reduced below 1.5 nm (four or five atomic layers) quantum tunnelling effects will lead to unacceptable leakage currents. Today's state of the art devices use layers less than 2 nm thick, and problems are expected within the next few years.

Even if problems of this kind are overcome there are still reasons to be dissatisfied with current computers, as certain interesting questions lie beyond the range of their abilities. It is well known that it is difficult to simulate the behaviour of large quantum mechanical systems as exact calculations require vast amounts of memory just to write down the Hamiltonian, and diagonalising such enormous matrices is a time consuming process. For example, the Hamiltonian describing a system of $n$ strongly coupled spin-1/2 nuclei is a $2^n$ by $2^n$ square matrix, and diagonalising this requires around $8^n$ operations. For this reason it is difficult to study systems with more than a dozen strongly coupled spins.

Behaviour of this kind is not confined to the simulation of physical systems: many types of mathematical problem are known which suffer from the same difficulty, that the time needed to solve a problem increases rapidly with the size of the problem being tackled. Indeed mathematical problems can be divided into several computational complexity classes according to how rapidly the solution time increases[14]. In particular problems can be divided into two main groups: tractable problems, which can be solved efficiently, and intractable problems which cannot[‡].

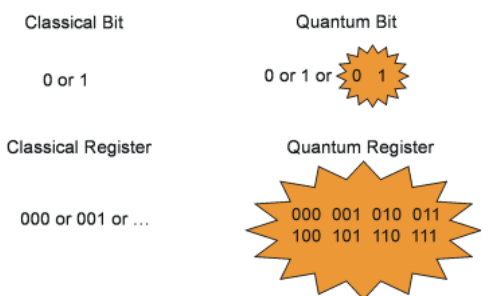

**Fig. 1** A classical bit is confined to two basic states, 0 and 1, but a quantum bit (qubit) can also enter superposition states, in which the qubit is effectively in both states at once. A classical register made up of $n$ bits can contain any one of $2^n$ possible numbers, while the corresponding quantum register can contain all $2^n$ numbers at the same time.

As discussed above, simulating quantum mechanics is an intractable problem, and yet quantum objects have no difficulty in obeying the laws of quantum mechanics. As any physical object can be considered as a "simulation" of some set of physical laws (for example, the motion of the planets around the sun can be considered as a simulation of Newton's laws), it seems that quantum mechanical objects have the ability to perform simulations which cannot be achieved with conventional classical computers. This was discussed by Feynman[15], who suggested that it might be possible to use one quantum system to simulate the behaviour of another, and that a well chosen system might be usable as a universal quantum simulator.

**Quantum information and classical information**

Classical information processing is concerned with bits, which are objects which can take two states, called 0 and 1. Bits can be implemented within quantum systems by simply assigning the two states to two eigenstates, |0⟩ and |1⟩; this approach is most natural for a two state quantum system, such as a spin-1/2 nucleus. Quantum bits, usually called qubits, are not, however, confined to these eigenstates, but can be found in superpositions (Fig. 1). Qubits in superposition states are in some sense in both basic states at the same time, and it is the ability of quantum systems to be found in superpositions, and more general states such as entangled states, which leads both to the difficulty in simulating quantum system and to the potential power of quantum computers.

Quantum information theory is an active and exciting area of modern physics, involving topics as diverse as the fundamental interpretation of quantum mechanics[16], the development of unbreakable cryptographic schemes[17], and the design of quantum algorithms which use quantum parallelism (Fig. 2) to tackle classically intractable problems[18]. There is also, of course, substantial interest in developing real quantum technologies, capable of implementing these ideas in physical systems, but while there was significant early success in using photons to demonstrate quantum communication schemes (such as quantum cryptography), it proved much more difficult to build even small quantum computers. The situation changed in 1996, however, when Cory *et al.* showed how NMR could be used[7].

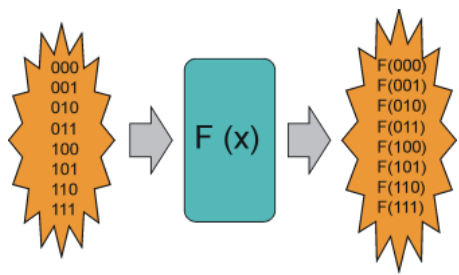

**Fig. 2** Quantum parallelism: a quantum processor can evaluate a function over a superposition of input states, effectively evaluating it over all possible inputs at the same time.

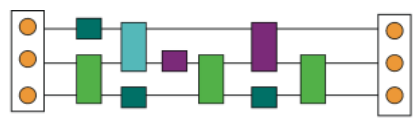

**Fig. 3** Any desired quantum information processing operation can be built as a network of single qubit and two qubit gates.

## NMR Quantum Computing

The basic requirements for experimental quantum computing are well known, and were famously listed by DiVincenzo[19]. There are five traditional requirements, but I shall consider requirements one and three together.

**Stable qubits**

In a real implementation of a quantum computer the two logical states of each qubit must be mapped onto the eigenstates of some suitable physical system; clearly the Zeeman levels of spin-1/2 nuclei in a magnetic field are a natural choice. There are five "obvious" candidate nuclei: $^1$H, $^{13}$C, $^{15}$N, $^{19}$F and $^{31}$P, and these have all been used, both alone and in various combinations, in experiments published to date; there are also many other more exotic candidates, such as $^{195}$Pt, which have not yet been used.

For a practical computer the qubits must have long relaxation times to prevent quantum effects from decohering away. While liquid state NMR relaxation times are not quite as long as one might wish, they are certainly long enough (around one second) for demonstration purposes.

**Quantum logic gates**

Coherent quantum processes, such as quantum algorithms, are described by unitary transformations. Fortunately, however, it is not necessary to implement arbitrary unitary transformations, as complex tranformations can be built up out of simpler building blocks. Just as classical information processing devices are constructed from networks of classical gates, quantum devices are built from networks of quantum logic gates (Fig. 3).

It is easy to show that any desired network can be built out of single qubit and two qubit gates[20]; thus only one spin and two spin interactions are necessary. NMR quantum computers use the scalar spin–spin coupling interaction ($J$ coupling) as their basic two qubit gate, with conventional multi-pulse NMR techniques used to sculpt the underlying Hamiltonian into the desired form.

**An initialisation scheme**

Before a quantum algorithm can be run it is necessary to set the computer to some well defined initial state. In most proposed designs this is realised by some sort of cooling process, allowing all the qubits to relax back to their energetic ground states, but this approach is not practical for NMR as the Zeeman energy splitting is small in comparison with the thermal energy ($kT$) at any reasonable temperature. This difficulty appeared to rule out NMR as a quantum computing technology; this conclusion is, however, over hasty. Although it is not possible with conventional NMR methods to produce a system in a pure ground state, it is possible to produce pseudo-pure states which act in much the same way.

**Readout**

The final requirement for practical computation is a method for measuring the state of one or more qubits so that the final result can be read out. Clearly this can be achieved by observing some sort of spectrum from the NMR system at the end of the calculation. Unfortunately this readout method has significant disadvantages which will be discussed below.

## A Historical View

The early history of NMR quantum computing can be found in a single remarkable publication[7] by Cory, Fahmy and Havel, in which most of the ideas developed over the next few years are described. The most important concept is that of the pseudo-pure state. A "conventional" quantum computer starts its calculations from the pure ground state, which for a two qubit device can be written as $|00\rangle$. This state is not normally accesible to NMR experiments, but as described by Cory *et al.* it is sufficient to form the state

$$(1-\varepsilon)\frac{\mathbf{1}}{4}+\varepsilon|00\rangle\langle00| \quad (1)$$

which behaves in a very similar fashion. In effect, the great majority of the NMR sample (in typical experiments $\varepsilon$ is around one part in $10^5$) can be ignored (Fig. 4), with only the small excess population of the ground state exhibiting any interesting behaviour[§].

In addition to describing the concept of pseudo-pure states, Cory *et al.* actually performed simple experiments using the two $^1$H nuclei in 2,3-dibromothiophene, demonstrating the concept of pseudo-pure states as well as their behaviour under simple quantum logic gates. At much the same time Gershenfeld and Chuang described a different procedure for generating "effective pure states" in NMR systems[21]; they did not, however, provide any experimental demonstrations of their scheme.

**The first algorithms**

While Cory *et al.* demonstrated the basic operations of quantum computing[7], it is debatable whether they actually implemented a quantum computation: it particular they did not run any actual algorithms. Several "toy" algorithms, which can be run on two qubit computers, have been known for many years and it was not long before the first implementations of these began to appear.

The first quantum algorithm was described by Deutsch[22–24], and this was also the first quantum algorithm to be implemented[¶]. Deutsch's algorithm is most accurately described in the language of binary function evaluation, where it provides a method for determining the parity of a function from one bit to one bit using only one function evaluation, but a more concrete picture can be obtained by thinking about coins. Normal coins have two sides, traditionally called heads and tails, while trick coins show the same pattern on both sides. To distinguish normal and trick coins it would seem necessary to examine both sides and compare the results, but Deutsch showed that with a quantum device the distinction could be made in a single glance.

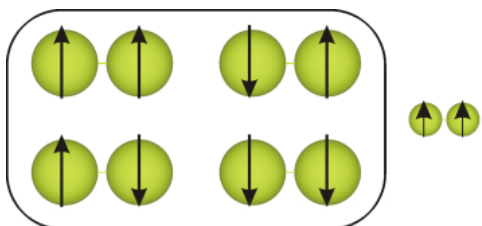

**Fig. 4** A pseudo-pure state of a two qubit system. There are nearly equal populations of the four spin eigenstates, and these do not contribute to the overall spectrum; only a signal from the small excess population is seen.

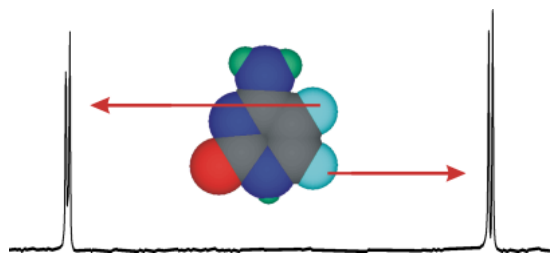

**Fig. 5** Deuterated cytosine (easily made by dissolving cytosine in D$_2$O) was used for the first NMR implementation of a quantum algorithm. The NMR spectrum contains two groups of lines, corresponding to the two nuclei as indicated. The doublet splitting arises from the *J* coupling, and is used to implement quantum logic gates.

This algorithm was first implemented[8] by NMR using the two $^1$H nuclei on a cytosine molecule dissolved in D$_2$O (Fig. 5), and this was rapidly followed by a second implementation[9] using the $^1$H and $^{13}$C nuclei in $^{13}$C labelled chloroform. These two systems were then used to implement a second toy algorithm, Grover's quantum search; this time the chloroform implementation[10] came first, with the cytosine implementation[11] coming a few months later.

**The golden age**

The three year period from early 1998 could be considered the "golden age" on NMR quantum computation, during which new results were demonstrated and new techniques were developed with startling rapidity. During this time many new algorithms were implemented, notably the Deutsch–Jozsa algorithm[26,27] (a generalisation of Deutsch's algorithm), quantum counting[28] (an extension of Grover's quantum search) and a simple example of order finding[29] (the basic element required for the famous quantum factoring algorithm[18] developed by Shor[**]). NMR quantum computers have also grown in size, with systems with three[26], four, five[27] and seven[30] qubits being described (Fig. 6). In addition to quantum algorithms these systems have also been used to demonstrate a wide range of quantum phenomena, such as quantum teleportation[††31], quantum error correction[32–34], and so called "Schrödinger Cat" states[‡‡30].

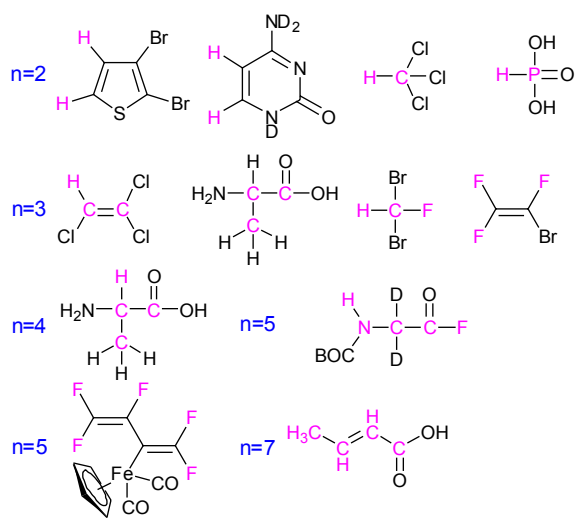

**Fig. 6** A selection of molecules that have been used to implement NMR quantum computing experiments. The number of qubits (n) is shown in blue, while the nuclei used as qubits are indicated in pink. Note that in the seven qubit system the three $^1$H nuclei in the methyl group are used as a single logical qubit.

## NMR Quantum Computing Today

After the hectic rush of the first few years, research into NMR quantum computing has settled into a period of steady progress. Most of the simple demonstrations which can be carried out with small numbers of qubits have now been performed, and the emphasis of current research has moved to more subtle issues.

**Building larger systems**

One major strand of research is, of course, the attempt to build larger NMR quantum computers. Although there are serious difficulties[35] in the way of building NMR devices large enough to be really interesting in their own right, it is certainly possible to make significant progress beyond the current point[36]. The ultimate limits to NMR quantum computing will be explored in the next section, but for the moment it suffices to note that most authors believe that the while ten qubit computers will be possible, systems with more than about twenty qubits will be very challenging. The current record is a seven qubit system[30], and it is likely that ten qubits will be achieved within a few years.

**Increasing spin polarisation**

As described previously, NMR quantum computing is implemented using pseudo-pure states[7] instead of pure states. These states can be characterised by their polarisation, $\varepsilon$: when $\varepsilon$ is equal to one the state is pure, while when $\varepsilon$ equals zero the state is said to be maximally mixed and cannot be used for computations. Clearly the state polarisation which can be achieved is limited by the underlying spin polarisation, which for conventional NMR systems arises from the Boltzmann distribution and is around one part in $10^5$. This tiny polarisation occurs because NMR operates in the high temperature regime, where the Zeeman energy gaps are much smaller than $kT$. Even worse, it is not possible to efficiently extract a large state polarisation from a high temperature spin polarisation: the fraction which can be extracted is roughly proportional to $2^{-n}$, where $n$ is the number of qubits in the system. Clearly this function falls off extremely rapidly as $n$ is increased.

This low state polarisation, and its unfortunate scaling behaviour, leads to two concerns. Firstly, the signal strength from an NMR quantum computer falls exponentially with the number of qubits, making large systems completely inpractical[37]. Secondly, the low state polarisation has given rise to more fundamental concerns[38] as to whether NMR quantum computing is real quantum computing at all! This question will be addressed in more detail below.

For both there reasons there is substantial interest in developing systems with much higher spin polarisations. This is also an important area of research for conventional NMR, and a number of approaches have been developed[35]. Unfortunately none of these are entirely appropriate for quantum computation, although one approach, based on *para*-hydrogen, has generated significant interest.

The obvious approach to generate high spin polarisations is to use some sort of optical pumping, but while this can be used to generate extremely high polarisations in noble gas nuclei (most notably $^3$He, $^{129}$Xe and $^{131}$Xe) such systems are of no use for quantum computation, which requires networks of coupled spins. It is, of course, possible to transfer the polarisation from the noble gases to some more interesting system, but so far the transfer efficiencies involved are far too low. There are many other techniques, such as CIDNP, which result in relatively high polarisations, but once again these are not high enough to be truly useful.

One important exception is the use of *para*-hydrogen induced polarisation (PHIP) in which the pure spin singlet state of the two $^1$H nuclei in the rotational ground state of a H$_2$ molecule is mantained during addition across a double bond[39–41]. In its simplest form this will only result in two polarised nuclei, but it should be possible to achieve extremely high polarisations, approaching a pure state. As discussed below, achieving a state polarisation of one third (the current record is about one tenth[41]) in a two qubit system would be a significant result, and this is an important target. Furthermore, it should be possible to extend the PHIP approach to larger spin systems by adding two or more H$_2$ molecules at the same time, effectively synthesising the quantum computer immediately prior to using it.

**Transferring NMR tricks to quantum computing**

One reason for the great speed with which NMR quantum computers were developed is the extraordinarily sophisticated level of quantum manipulation used in everyday conventional NMR experiments. Decades of experience manipulating spin systems has produced an extensive library of techniques and tricks, many of which can be transferred to NMR quantum computing, and from there into other quantum computing technologies.

One example of this process is the design of quantum logic gates. Early research on NMR quantum computing sought to implement logic gates, such as the controlled-NOT gate, which are widely used in theoretical models of quantum computing, but while it is fairly simple to do this gates of this kind are not the most natural gates to use in NMR. More recently authors have begun to describe these gates in a more natural fashion, leading to simpler implementations[12]. Other authors have drawn analogies between quantum logic gates and more conventional NMR pulse sequences, such as spin-state selective excitations[42], and this has led to more efficient ways of implementing complex multi-qubit gates[43].

Turning from theory to experiment, one NMR trick which is likely to play a role in other quantum computing technologies is the use of composite rotations, also known as composite pulses[44]. Any unitary transformation can be described as a rotation, and single qubit gates correspond naturally to rotations on the Bloch sphere. While these are easy to describe, they are not quite so easy to perform, as experimental errors and non-idealities inevitably lead to incorrect rotation angles and axes. These effects are well known in NMR, where the cummulative effects of errors in complex multi-pulse sequences can be a serious problem. Composite pulses, in which a single RF pulse is replaced by a sequence of pulse performing the same overall process but with greater error tolerance, provide a simple and effective solution. Unfortunately most of the composite rotations developed for conventional NMR experiments cannot be used for NMR quantum computation, as they only work well in special cases. The basic idea, however, remains sound, and composite pulse sequences which can be used for quantum computing have been developed and demonstrated[45]. These ideas are likely to find application is any eventual implementation of quantum computing.

**Transferring quantum computing tricks to NMR**

While there has been some success in transferring techniques from conventional NMR into quantum computing, there is also interest in the reverse approach, seeking ideas from quantum information theory which can be applied in more conventional NMR experiments. To date, this approach has not proved particularly useful: while ideas from quantum information theory have provided some insights into why current NMR pulse sequences work as they do[46], they have not yet been used to invent important new pulse sequences. One minor new development which has come from NMR quantum computing is a method for deriving efficient refocussing schemes[47], but this is yet to find any wider use.

## The Limits to NMR Quantum Computing

As mentioned above, there appear to be quite strong limits on the size of NMR quantum computers which can be constructed. These limits arise from several quite distinct difficulties, and so even if one problem is overcome it will remain difficult to make progress beyond a certain limit[§§]. Furthermore, some of the limitations seem to be inherent in liquid state NMR, and are thus particularly difficult to overcome. The major factors limiting the development of large systems[35] are discussed below.

### Polarisation

One obvious difficulty in constructing large NMR quantum computers is the way in which the available state polarisation falls off as the number of qubits increases. While the exact polarisation achievable depends on experimental details an upper limit can be estimated from the population difference between the lowest and highest energy states; for a homonuclear spin system with $n$ qubits this is given by[37]

$$\varepsilon = \frac{2\sinh(nh\nu/2kT)}{2^n \cosh^n(h\nu/2kT)} \qquad (2)$$

where $h\nu$ is the Zeeman splitting and $kT$ is the thermal energy. In the high temperature limit ($h\nu \ll kT$) this simplifies to

$$\varepsilon \approx \frac{nh\nu/kT}{2^n} \qquad (3)$$

and so the available polarisation decreases exponentially with the size of the system. If the size of the NMR sample is held constant then the signal strength will fall off exponentially with the number of qubits; alternatively to keep a constant signal size it would be necessary to use exponentially large samples.

This exponential fall off is often discussed as if it were the most important limitation on the implementation of NMR quantum computers, but this is a great oversimplification. This problem is not an intrinsic feature of NMR: rather it arises because current implementations use thermal ensembles in the high temperature limit. Clearly this could in principle be overcome by reducing the effective spin temperature, or (more realistically) by generating a non-Boltzmann pattern of spin populations, as discussed above.

### Decoherence

In order to perform large quantum computations it is necessary to ensure that the decoherence time is long in comparison with the time required to implement a quantum logic gate, so that it is possible to implement many logic gates before any significant decoherence occurs[48]. NMR systems have fairly long relaxation times, and thus it might be expected that decoherence would not be a major problem. Unfortunately decoherence processes do not just affect single spins, but can also apply to multi-spin coherent states (multiple quantum coherence), and such states can have quite short relaxation times, even when the single spin relaxation times ($T_1$ and $T_2$) are fairly long.

More seriously, even when NMR decoherence times are long the time required to implement quantum logic gates is also quite long, and so the critical ratio is not as large as one might hope. For two directly coupled spins the time required for a two qubit logic gate is about half the inverse of the corresponding $J$ coupling, and since useful $J$ couplings lie in the range of about 5–1000 Hz two qubit gates can take as long as 100 ms to implement.

### Selective Excitation

The problem of selective excitation is rarely discussed, but is in fact likely to be one of the most severe limitations on the development of NMR quantum computers. All computers (quantum or classical) require some means by which individual bits or qubits can be addressed. This is usually achieved by spatial localisation, that is using the fact that different qubits are found in different locations in space. This approach is not possible for NMR quantum computers, as qubits are implemented using an ensemble of spins, dispersed throughout the sample and in constant rapid motion. Instead it is necessary to use frequency selection, picking out each nucleus according to its characteristic Lamor precession frequency. Unlike physical space, the bandwidth available for frequency selection is predetermined and quite limited (less than 10 kHz for $^1$H nuclei).

### Reinitialisation and error correction

More complex computations require the use of larger numbers of logic gates, and it was originally thought that this effect would limit the size of any quantum computer, as the time needed for a computation would inevitably begin to exceed the decoherence time. The remarkable discovery of quantum error correction techniques and fault tolerant quantum computation[48–51] has (in principle) allowed this problem to be sidestepped. In essence it has been shown that as long as the ratio of the gate time to the decoherence time can be reduced below some critical threshold then it is possible to use error correction methods (adapted from classical error correcting codes) to completely remove the effects of decoherence.

Quantum error correction techniques are likely to play a central role in any large scale implementation of quantum computing, but while the basic approach has been demonstrated using NMR techniques[32–34] it is not possible to perform effective error correction within current NMR implementations. This is because effective error correction requires the ability to selectively reset certain qubits to zero in the middle of a computation, while current NMR methods only allow the states of qubits to be reset once, at the start of the computation.

More fundamentally these difficulties arise because of the NMR readout scheme: with NMR quantum computers readout is based on making weak measurements on ensembles of spins, rather than making strong (projective) measurements on single spins. Projective measurements are extremely useful, as they provide an intrinsic resetting mechanism, and their absence in NMR implementations is probably the most fundamental weakness of the approach[35].

## The Question of Entanglement

The low polarisation states used for NMR quantum computing have important practical consequences, but have also led to claims[38] that the NMR approach is not proper quantum computing at all! These claims rest on technical arguments about the nature and importance of entanglement, a property of quantum mechanical systems where the behaviours of parts of a composite system appear to share a mysterious link. While understanding entanglement can raise philosophical problems, its properties in simple situations are well understood, and the creation and manipulation of entangled states plays a central role in quantum information processing. It has long been believed that the increased power available to quantum computers is linked to their use of entanglement[52], although this has never been completely proved.

Conventional models of quantum computing assume that the quantum computer exists in a single pure quantum state, while NMR implementations are based on highly mixed states deriving from thermal ensembles. The nature of entanglement in such mixed states is more complex than in pure states, and is not quite so well understood. In essence the problem arises because there is no unique way to decompose a mixed state as a mixture of pure states[¶], and so while some mixed states may appear to include an entangled component in the mixture, it may be possible to describe the mixture without using entangled states[12,38]. Mixed states of this kind are not provably entangled by physical tests, and there is considerable doubt as to whether any of the characteristic properties of entangled systems can be used.

It is known that with a pseudo-pure state of a two qubit system it is not possible to prepare a provably entangled state unless the polarisation lies above the critical threshold[53] of 1/3; for this reason there is considerable interest in using *para*-hydrogen techniques to prepare pseudo-pure NMR states above this limit. Similar results have been obtained for systems with larger numbers of qubits[38], and no NMR system demonstrated so far has involved physically provable entanglement. Furthermore, it has been shown that current NMR systems cannot be used for an efficient implementation of Shor's factoring algorithm[54].

Despite these apparently damming conclusions, the situation is far from clear, however. Attempts to describe NMR quantum computing experiments with purely classical models[55] have not been successful, and it has also been suggested that NMR methods might be suitable for implementing simulations of other quantum systems[56]. No matter what the eventual outcome is, studying NMR has certainly had major effects on our understanding of entanglement and its uses.

## Other Quantum Computing Technologies

NMR is certainly not the only technology which has been studied as a method for building quantum computers: before the first NMR experiments were conducted there was substantial research into the use of trapped ions, and more recently there has been great interest in the possibilities offered by solid state systems. Within the last year a remarkable proposal has also been made for an "all optical" implementation.

### Trapped ions

It has long been known that it is possible to trap ions in free space using electric and magnetic fields, and then to slow them down using a variety of laser cooling methods. Such trapped ions are useful for a range of purposes, including the possibility of quantum computation[57], using pairs of electronic levels as qubits. Laser pulses can be used to induce transitions between energy levels, and thus implement single qubit gates, while the strong Coulomb forces between pairs of ions provide a two qubit interaction, and thus a route to arbitrary quantum logic gates. The large energy gaps associated with electronic transitions means that it is easy both to set the initial states of individual ions and to observe their states at the end of a computation.

From the description above it might seem that trapped ions would provide an ideal implementation of quantum computing, but while the basic elements of quantum computation have been demonstrated actually implementing a complete quantum algorithm has proved extremely difficult. Even when this is achieved it will be difficult to use the trapped ion approach to build large quantum computers. However the fact that, unlike NMR quantum computers, ion trap implementations use single quantum objects (rather than ensembles) for their qubits means that even a small ion trap quantum computer would be of great interest to researchers.

### Solid state systems

While the two most successful implementations to date have used liquid and gas phase technologies it is widely agreed that if a general purpose quantum computer ever enters widespread use it will almost certainly be based on a solid state approach. There are many reasons behind this widespread belief, but perhaps the simplest is the fact that objects in solid state devices stay where they are, rather than moving around! Many of the difficulties in scaling up NMR quantum computers can be linked to the lack of spatial localisation, while a quite inordinate amount of effort is spent in ion trap implementations on simply keeping the individual ions still. Other obvious advantages include the enormous technical sophistication of solid state fabrication processes, and the low marginal cost of manufacture once an initial device has been made.

Many different approaches have been suggested for using solid state technologies, and it is too early to make any firm comments on their relative merits. There are, perhaps, three leading schemes, based on single spin ENDOR[58] (a combination of NMR and ESR), quantum dots[59], and Josephson junctions[60].

### Optical systems

While light is often thought of in a fairly classical manner, photons are of course quantum objects and might in principle be useful for quantum information processing. In fact photons are extremely well suited to certain simple quantum communication schemes, such as quantum cryptography. They are not, however, well suited to more complex tasks, such as quantum computing, as the interaction between pairs of photons is normally extremely weak. One solution to this problem is to combine photons with atoms or ions to create hybrid systems[57]. Recently, however, a new scheme has been proposed which should permit an all optical implementation of quantum computing[61]. This scheme requires some basic elements (notably reliable single photon sources and single photon detectors) which are not currently available, but methods for implementing such devices (which would be useful for many other purposes) are being sought.

## Summary

Quantum information processing offers the possibility of going beyond the limitations apparently imposed by classical physics. In particular quantum computation may allow us to tackle problems well beyond the abilities of any classical computer. If these hopes are to be realised, however, we will have to develop suitable quantum technologies.

In the last few years there have been huge advances in constructing small quantum computers using techniques based on conventional liquid state NMR experiments. While these computers are too small for any practical use their development has generated significant experimental interest in a field long dominated by theory. Most of the basic techniques of quantum information processing have now been demonstrated, as have a number of simple quantum algorithms.

Although building small NMR devices has proved surprisingly simple, there are formidable difficulties in scaling these systems up to. Many of these problems are being tackled, and it is likely that some of them will be solved, but most authors believe that it will not be possible to build NMR quantum computers large enough to solve real computational problems.

A wide range of other approaches have been suggested for building large quantum computers, most notably schemes based on solid state physics, but although some of these techniques show promise, it is not yet clear which (if any) will eventually succeed. In the meantime it is likely that NMR will remain the leading technology for many years to come. Furthermore, the development of small NMR quantum computers has provided many useful insights into methods for building larger systems, and into the nature of quantum computing itself.

## Footnotes

\* Much of the early research into quantum information processing was conducted within a small community of physicists and mathematicians and communicated in a relatively informal manner. The Los Alamos National Laboratory (LANL) e-print archive has long played a central role: some manuscripts are never formally published, but simply made available at http://arxiv.org/archive/quant-ph, and many conventional papers can be found there long before formal publication. Websites managed by individual research centres can also be useful, and the Oxford Centre for Quantum Computation (http://www.qubit.org/) is a good place to start.

† This empirical observation is usually ascribed to Gordon Moore, the founder of Intel, although the form in common use today is a summary of various observations made by him and by others.

‡ This discussion implicitly assumes the so called "Church–Turing thesis" which states that the computational complexity of a problem is essentially the same (technically, varies only polynomially) on any physically reasonable design of computer. The discovery of quantum computing appears to overthrow this long held dogma.

§ This description is viewed by theoreticians with suspicion, and is sometimes referred to as "the preferred ensemble fallacy". It does, however, provide a simple and broadly correct view of the situation.

¶ In fact the algorithm actually implemented was a later version developed by Cleve *et al.*, but this later version is commonly referred to as Deutsch's algorithm.

\*\* Shor's quantum factoring algorithm has great importance, as the security of current public key cryptography schemes rests on the presumed difficulty of factoring; the construction of a quantum computer capable of fully implementing Shor's algorithm would immediately render the current public key infrastructure obsolete.

†† Quantum teleportation is the name given to a process in which the state of one quantum system is transferred to another remote system without any knowledge of the form of the state being transferred. In NMR quantum teleportation the state of one nuclear spin is transferred to another nuclear spin a few Ångströms away; clearly this is much less satisfactory than corresponding optical experiments in which the state of a photon can be teleported over about one metre.

‡‡ Schrödinger's Cat is the name given to situations where quantum mechanics indicates that the state of a macroscopic object is entangled with the state of a microscopic quantum system. Within quantum information processing the name is sometimes also used loosely to refer to entangled superpositions of four or more microscopic particles. Many physicists are unhappy with this development, and the name "Schrödinger Kitten" has been suggested as an alternative, less misleading, description.

§§ The clear limitations on the development of large NMR quantum computers are frequently made much of by proponents of alternative quantum technologies. It should, however, be remembered that one of the reasons why the limits to NMR are so clear is because NMR is already such a well developed system. By contrast, many other quantum technologies are little more than paper proposals, and it is by no means clear what difficulties will surface in any real implementation.

¶¶ This lack of a unique decomposition for a mixed state lies at the heart of theoretical suspicions about the conventional model of pseudo-pure states depicted in Fig. 4.